\documentclass[12pt]{article}
\usepackage{graphicx}

\begin{document}
\begin{titlepage}
%\maketitle
\begin{center}
\bf{Transport properties of the Hubbard model in low dimensions}
\end{center}
\begin{center}
V.Celebonovic
\end{center}
\begin{center}
Institute of Physics,Pregrevica 118,11080 Zemun-Belgrade,Serbia
\end{center}

\begin{center}
e-mail: vladan@ipb.ac.rs 
\end{center}
 
published in: J. of Optoelectronics and Advanced Materials-

Symposia,$\bf{1}$,pp.494-497 (2009).

%\end{centerline}
%\address{$^2$Instituto de Astrof\'isica de Canarias, 38200 La Laguna, Tenerife, Spain}
%\Email{emg}{ll.iac}{es}

% Running titles
%\markboth{THE O III NARROW LINE SHAPES IN SPECTRA OF Mrk 817}{L. \v C. POPOVI\'C and E. MEDIAVILLA}

% Abstract
\begin{abstract}
	The Hubbard model describes interacting electrons on a lattice,a situation which occurs in many solid state materials and devices.The aim of the present paper is to briefly discuss this model and its applications in the study of transport properties of quasi-one dimensional and quasi-two dimensional systems. Several examples are taken from materials and fields intensively studied in recent years: organic conductors and semiconducting heterostructures. 
\end{abstract}
\end{titlepage} 

% Section and subsection
\section{Introduction }
All natural sciences are devoted to an easily formulated aim: trying to discover,explain and predict how natural phenomena occur. Performing experiments on any kind of a specimen in any natural science means putting the system under study into interaction with some kind of external probe,and then measuring the system's response. If this interaction is weak,the system's response is linear and in solid state physics this is usually called the one electron picture. 

It has been shown in the last few decades that there exist materials which can not be described within the one-electron picture. Examples of such materials are the organic conductors and the high-temperature superconductors. Quasi-one-dimensional (Q1D) organic conductors were synthetized in the last century [1]. Their general chemical formula is $(TMTSF)_{2}$${X}$ where $(TMTSF)_{2}$ is a complicated compound called bi-tetra-methyl-tio-seleno-fulvalene,and $X$ is any anion which can be added to the central complex. Some examples of the anions which can be added are: $ClO_{4}$,$FSO_{3}$,$ReO_{4}$..For a recent review and some history of the field see [2]. 

Early experiments have already shown that the electrical conductivity of these new materials,which later became known as the Bechgaard salts,{\it can not} be described within the standard theory of metallic conductivity [3]. This discrepancy was explained by the fact that the electrons in Bechgaard salts are correlated. 

The simplest theoretical model applicable to the Bechgaard salts is the Hubbard model (HM) [4]. The present paper is devoted to a brief introduction to this model and its transport properties, and applications of these theoretical results to two kinds of systems,under the limitation of low dimensionality. This limitation is motivated both mathematically and physically.Mathematically,the Hubbard model has so far been solved only for one dimensional systems (1D) [5]. Physically,it has been shown in recent years that low dimensional materials have extraordinary properties.

%\end{titlepage}

\section{The Hubbard model }

Around the middle of the last century Mott developed a model of the metal to insulator transition (MIT),and showed what influence correlations can have on the results of the one electron picture [6]. The developement of the HM was motivated by the need to develop a microscopic model of the Mott transition.  

The starting ingredient of the HM is a collection of atomic orbitals.The model is developed under the assumption of tight binding. This means that an electron's wave function is assumed to be centered on one atomic site,and that an electron can "hop" just one lattice spacing at a time. The rigorous formulation of the HM turned out to be too complex for any practical applicability,so Hubbard had to introduce considerable simplifications. Basically,the Hamiltonian is given as a sum of a "free" kinetic term $H_{0}$ and the interaction term $H_{I}$. It can be shown that the simplest expression for the Hubbard Hamiltonian in one spatial dimension is: 

\begin{equation}
	H=-t\sum_{i,\sigma}(c_{i+1,\sigma}^{+}c_{i,\sigma}+c_{i,\sigma}^{+}c_{i+1,\sigma})+U\sum_{i}n_{i,\uparrow}n_{i,\downarrow}
\end{equation}

where $t$ is the hopping integral,$U$ the on-site repulsion,and all the other symbols have their usual second quantisation meanings. 

To study the interplay of interactions and disorder in electronic systems disorder can be introduced into the Hubbard model.The most common type of disorder is that of a random potential; for the Hubard model this translates into random site energies.An equivalent formulation is that of random bonds,which practically means random hopping parameters.If we allow for a paralel magnetic field which couples to the electron spin,the Hamiltonian for 2D calculations becomes 
\begin{equation}
	H=-\sum_{i,j,\sigma}t_{ij}c_{i \sigma}^{+}c_{j \sigma}+U\sum_{i}n_{i,\uparrow}n_{i,\downarrow}-\sum_{j,\sigma}(\mu-\sigma B)n_{j \sigma}
\end{equation}
$B$ denotes the magnetic field which couples to the electron spin,and $\mu$ is the chemical potential. The strength of the disorder is tuned by taking the hopping parameters $t_{i j}$ from a probability distribution of the form $P(t_{i j})=1/\Delta_{t}$ for $t_{i j}\in[t-\Delta_{t}/2,t+\Delta_{t}/2]$ and zero otherwise.

\section{The transport properties} 

Studies of the conductivity of the Bechgaard salts are of interest only under low temperature conditions,which means that electron-electron scattering is the dominant scattering mechanism. The electrons in these materials can be described as a Fermi (or Luttinger) liquid.Accordingly,it should be possible to apply methods of the theory of fluids to the calculation of the transport properties of the Bechgaard salts.Various methods for the calculation of the conductivity,assuming that the Hamiltonian is known,have recently been reviewed in [7]. In work on the Bechgaard salts,the "memory function" method was used [7].
\newpage 
This method was developed in the seventies,and within it the conductivity can be calculated as follows: 
  
\begin{eqnarray}
	\chi_{AB}(z)=<<A;B>>= -i\int_{0}^{\infty}\exp{izt}<[A(t),B(0)]> dt
\end{eqnarray} 

where $A=B=[j,H]$,$j$ denotes the current operator, and 

\begin{equation}
	\sigma(z) = i\frac{\omega_{P}^{2}}{4\pi z} [1-\frac{\chi_{z}}{\chi_{0}}]
\end{equation}
 
\section{Results}

\subsection{The Bechgaard salts}

The electrical conductivity of the Bechgaard salts was calculated using Eqs.

(1),(3) and (4). Full details of the calculation have been discussed at length in [8]. It was assumed there that the electrons in a Bechgaard salt can be approximated as a Fermi liquid (FL). Field theory shows that the FL model fails in 1D systems (for example [10]). However,the Bechgaard salts are experimentally known to be quasi one-dimensional,although recent experiments suggest that a 2D Fermi surface exists in some of them [11], which can be taken as a justification for the applicability of the FL model.

The first step in the calculation was to determine the chemical potential of the FL on a lattice. The resulting expression has the form of a quotient,and fits to the known result of Lieb and Wu [5]. The next step was the determination of the susceptibility $\chi$,and the final one was the calculation of the conductivity. Various mathematical approximations were made,and they are discussed in [8]. 

The following figure represents an example of the temperature dependence of the conductivity of the Bechgaard salts. The curve is normalized to unity at $T=116K$ and it is drawn for two different values of the band filling $n$. Experimentally, varying the band filling means doping the material with an electron donor or an acceptor.The calculation was performed without taking into account the influence of high external pressure,which is usually varied in experiments (for example [9]). 
\newpage    
%\vfil\newpage

% Table
%\caption{Calculated positions for the next five years}
%\end{center}
%\end{table}

% Figure (in PS or EPS format)
\begin{center}
\begin{figure}
\includegraphics[width=9cm,height=9cm]{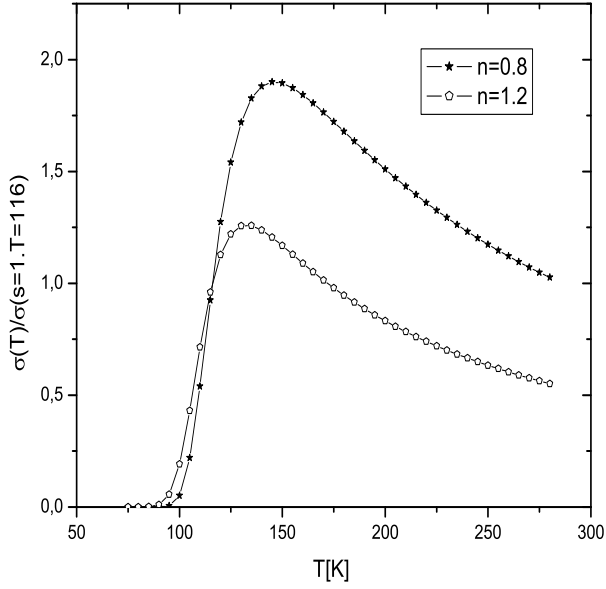}
\caption{\it{Normalzed conductivity of a Bechgaard salt for two values of the band filling} }
%\end{center}
\end{figure}
\end{center}

Later work has shown that a suitable change of variables gives the possibility of taking into account the influence of high external pressure on the electrical conductivity [12]. The real part of the susceptibility has the form
\begin{equation}
	\chi_{R}(\omega)=\sum_{i}\frac{A_{i}}{\omega+q_{i}t}
\end{equation}

where $q_{i}$ are numerical constants. The functions $A_{i}$ contain all the parameters of the problem which means also the value of the lattice constant,except the frequency. Starting from this result,one can calculate the immaginary part of the susceptibility,and after that the real part of the conductivity. The final result for the real part of the conductivity has the following form

\begin{equation}
\sigma_{R}(\omega_{0})=\frac{1}{\chi_{0}}\sum_{i}\frac{A_{i}}{(q_{i}t)^{2}}\frac{\ln[(\frac{\omega_{0}}{q_{i}t})^{2}]}{1-(\frac{\omega_{0}}{q_{i}t})^{2}}
\end{equation}
As the equation of state of the Bechgaard salts is not known, the influence of high pressure on the conductivity is taken into account through the change of the lattice constant. 

Apart the electrical conductivity,experiments in the past $\approx10$ years have given the possibility of measuring the thermal conductivity of the organic conductors. First experiments were made on the Bechgaard salts,but these were quickly followed by studies of quasi two dimensional organic conductors. Examples are [16],17]. 

\subsection{Two dimensional systems}   
The basic numerical methos used in applying the Hubbard model to two dimensional systems is the so called Determinant Quantum Monte Carlo Method (DQMCM),developed at the University of Santa Barbara.A brief review of the method is presented in [13] and references given there. Details on the calculation of the partition funaction are also avaliable there. 
%\begin{center}
%\begin{figure}
%\includegraphics[width=10cm,height=8cm]{fig2.eps}
%\caption{\it{Temperature dependence of the conductivity $\sigma_{DC}$ for various values of the disorder $\Delta$ at $U=4$ for $<n>=0.5$.Calculations were performed on an $8x8$ lattice,and data points are averages over 4 realizations of a given disorder.}}
%\end{center}
%\end{figure}
%\end{center}
Interesting results have been obtained in applying DQMCM to the 2D Hubbard model [14]. Plots of the conductivity  as a function of the temperature $T$ for various values of the disorder $\Delta_{t}$ show an indication of a metal to insulator transition above a critical disorder of $\Delta_{c}\cong2.7$. If such a behaviour persists to very low temperatures,and in the thermodynamical lmit, it would describe a ground state metal to insulator transition driven by disorder.The temperature is expressed in units of $t$. The parameter varied in laboratory experiments is the electron density, while in theoretical work on the Hubbard model one varies the disorder strength $\Delta$. It can be shown that the two approaches are equivalent.

It has been shown [15] that a Zeman magnetic field reduces the conductivity of a conducting disordered electronic 2D system with a fixed disorder and a variable field, and it becomes temperature independent after a certain critical value of the field. At larger field strenghts there are indications of insulating behaviour.The occurence of a metal to insulator transition driven by a magnetic field is in agreement with real laboratory experiments such as [18]. In that particular experiment,the value of the resistivity at the transition was found to be density dependent,which implies that it can be controlled by high external pressure. It is interesting to note that a recent study of a 2D Hubbard model with site disorder gives physically similar results [13]. 
\newpage  
%\begin{center}
%\begin{figure}
%\includegraphics[width=8cm,height=8cm]{sigU4nv.eps}
%\caption{\it{Temperature dependence of the conductivity $\sigma_{DC}$ for various values of the disorder $\Delta$ at $U=4$ for $<n>=0.5$.Calculations were performed on an $8x8$ lattice,and data points are averages over 4 realizations of a given disorder.}}
%\end{center}
%\end{figure}
%\end{center}
%It has been shown [15] that a Zeman magnetic field reduces the conductivity of a conducting disordered electronic 2D system with a fixed disorder and a variable field, and it becomes temperature independent after a certain critical value of the field. At larger field strenghts there are indications of insulating behaviour.The occurence of a metal to insulator transition driven by a magnetic field is in agreement with real laboratory experiments such as [18]. In that particular experiment,the value of the resistivity at the transition was found to be density dependent,which implies that it can be controlled by high external pressure. It is interesting to note that a recent study of a 2D Hubbard model with site disorder gives physically similar results [13]. 

\section{Conclusions}

In this paper we have briefly discussed the Hubbard model and some of its applications to the study of transport properties of quasi-one-dimensional and quasi-two-dimensional systems. Due to space limitations,many interesting aspects have not been discussed. In the 1D case,the Hubbard model was apllied to the Bechgaard salts,at first without but then with taking into account effects of external pressure. 

Concerning possible metal to insulator transitions in 2D electronic disorder systems,it has been shown that interactions enhanec the conductivity. At low temperature,a transition from a metallic state at weak disorder to an insulating state at strong disorder can occur. 

In both dimensionalities,it would be interesting infuture work to  allow for longer range interactions. It seems that a necessary step in applications of the Hubbard model is to investigate in detail its thermodynamcal properties like the compressibility or susceptibility. 

\section{Acknowledgement} 

I am grateful to dr.P.J.H.Denteneer from the Lorentz Institute of Leiden University in Holland for discussions and correspondence on various aspects of the Hubbard model. The preparation of this work was financed by the Ministry of Science and Technology of Serbia under its project 141007. 
  
% References

%\section{References}
{}

%Hewitt, A., Burbrdge, G. : 1989, \journal{Astrophys. J. Suppl. Series}, \vol{75}, 297.

%Mediavilla, E., Insertis, F. M. : 1989, \journal{Astron. Astrophys.} \vol{214}, 79. 

%Netzer, H. : 1990, \journal{Active Galactic Nuclei, eds. R. D. Blandford, H. Netzer \& L. Woltjer}, Saas-Fee Advanced Course 20, Berlin: Springer -- Verlag.  

%Osterbrock, D. E. : 1989, \journal{Astrophysics of Gaseous Nebulae and Active Galactic Nuclei}, Mill Valley, California. 
%\endreferences


\begin{thebibliography}{00}

\bibitem{[1]}
K.Bechgaard, C.S.Jacobsen,K.Mortensen et al.,Solid State 

Comm.,$\bf{33}$,p.1119 (1980). 


\bibitem{[2]}
D.J\'erome,Chem.Rev.,$\bf{104}$,p.5565,2004. 

\bibitem{[3]}
D.J\'erome and F.Creuzet in $\it{Novel}$ ${Superonductivity}$,ed.by S.A.Wolf and V.Z.Kresin,
p.103,Plenum Press,London,(1987).

\bibitem{[4]}
J.Hubbard,Proc.Roy.Soc.,$\bf{A276}$,238 (1963).

\bibitem{[5]} 
E.H.Lieb and F.Y.Wu,Physica $\bf{A321}$,1 (2003).

\bibitem{[6]}
N.F.Mott, Rev.Mod.Phys.,$\bf{40}$,677 (1968).

\bibitem{[7]}
U.Balucani,M.H.Lee,V.Tognetti,Phys.Rep.,$\bf{373}$,409 (2003).

\bibitem{[8]} 
V.Celebonovic in $\it{Trends}$ ${in}$ ${Material}$ ${Science}$ (ed.by B.M.Caruta), Nova Science Publishers,N.Y.,p.241,(2006),and preprint 

cond-mat/0412146.

\bibitem{[9]}
S.Kagoshima, R.Kondo,N.Matsushita et al.,J.Low Temp.Phys.,$\bf{142}$,413 (2006).  

\bibitem{[10]} 
T.Giamarchi,preprint arXiv cond-mat/0702565 (2007). 

\bibitem{[11]} 
W.Kang,Y.J.Jo,H.Kang et al.,Synth.Met.,$\bf{137}$,1189 (2003).

\bibitem{[12]} 
V.Celebonovic,2009,in preparation.

\bibitem{[13]}
P.B.Chakraborty, P.J.H.Denteneer,R.Scalettar, Phys.Rev.,$\bf{B75}$,125117 (2007). 

\bibitem{[14]} 
P.J.H.Denteneer,R.Scalettar,N.Trivedi,Phys.Rev.Lett.,$\bf{83}$,4610  (1999).


\bibitem{[15]} 
P.J.H.Denteneer and R.Scalettar, Phys.Rev.Lett.,$\bf{90}$,246401 (2003).

\bibitem{[16]}
S.Belin and K.Behnia,Phys.Rev.Lett.,$\bf{79}$,2125 (1997). 

\bibitem{[17]}
J.Wosnitza.S.Wanka,J.Hagel et al.,Synth.Met.,$\bf{133-134}$,201 (2003). 

\bibitem{[18]} 
J.Yoon,C.C.Li,D.Shahar et al.,Phys.Rev.Lett.,$\bf{84}$, 4421 (2000).
\end{thebibliography}
\end{document}